\documentclass[final]{svjour3}
\usepackage{graphicx}
\usepackage{rotating}
\usepackage{amssymb}
\usepackage{amsmath}
\usepackage{mathptmx}
\usepackage[square,numbers]{natbib}
\usepackage[utf8]{inputenc}
\usepackage{subfig}
\usepackage{color}
\usepackage{array}
\usepackage[color=red!20]{todonotes}
\usepackage{multirow}
\usepackage{sidecap}
\makeatletter
\journalname{Journal of Low Temperature Physics}

\bibpunct{[}{]}{,}{n}{}{,}

\begin{document}

\def\fullwidth{0.95}                
\def\halfwidth{0.45}                
\def\thirdwidth{0.25}				

\newcommand{\hdblarrow}{H\makebox[0.9ex][l]{$\downdownarrows$}-}
\def\starfire{{\sc starfire}}
\def\CII{[C{\sc ii}]}
\def\NII{[N{\sc ii}]}
\newcommand{\mum}{\ensuremath{\mu\mathrm{m}}}
\newcommand{\nepunits}{W\,Hz$^{-1/2}$}
\newcommand{\lt}{<}
\newcommand{\gt}{>}

\newcommand\apj{ApJ}
\newcommand\aap{A\&A}
\newcommand\procspie{Society of Photo-Optical Instrumentation Engineers (SPIE) Conference Series}
\newcommand\prb{Phys. Rev. B}

\title{Development of Aluminum LEKIDs for Balloon-Borne Far-IR Spectroscopy}

\author{S. Hailey-Dunsheath$^1$ \and A.~C.~M.~Barlis$^2$ \and J.~E.~Aguirre$^2$ \and C.~ M.~Bradford$^3$ \and J. G. Redford$^1$ \and T.~S.~Billings$^2$ \and H. G. LeDuc$^3$ \and C.~M.~McKenney$^4$ \and M.~I.~Hollister$^5$ }

\institute{Department of Physics, Name University,\\ City, STATE zip, Country\\ Tel.:\\ Fax:\\
\email{name@email.com}}

\institute{1: California Institute of Technology, Mail Code 301-17, 1200 E. California Blvd., Pasadena, CA 91125, USA; 
\email{haileyds@caltech.edu} \\
2: University of Pennsylvania Department of Physics \& Astronomy, 209 S 33rd St., Philadelphia, PA, 19104, USA \\
3: Jet Propulsion Laboratory, 4800 Oak Grove Drive, Pasadena, CA, 91109, USA \\
4: National Institute of Standards and Technology, 325 Broadway, Boulder, CO, 80305\\
5: Fermi National Accelerator Laboratory, PO Box 500, Batavia IL 60510
}

\maketitle

\begin{abstract}

We are developing lumped-element kinetic inductance detectors (LEKIDs) designed to achieve background-limited sensitivity for far-infrared (FIR) spectroscopy on a stratospheric balloon. The Spectroscopic Terahertz Airborne Receiver for Far-InfraRed Exploration (STARFIRE) will study the evolution of dusty galaxies with observations of the [CII] 158 \mum\ and other atomic fine-structure transitions at $z=0.5-1.5$, both through direct observations of individual luminous infrared galaxies, and in blind surveys using the technique of line intensity mapping. The spectrometer will require large format ($\sim$1800 detectors) arrays of dual-polarization sensitive detectors with NEPs of $1 \times 10^{-17}$ \nepunits. The low-volume LEKIDs are fabricated with a single layer of aluminum (20 nm thick) deposited on a crystalline silicon wafer, with resonance frequencies of $100-250$ MHz. The inductor is a single meander with a linewidth of 0.4 \mum, patterned in a grid to absorb optical power in both polarizations. The meander is coupled to a circular waveguide, fed by a conical feedhorn. Initial testing of a small array prototype has demonstrated good yield, and a median NEP of $4 \times 10^{-18}$ \nepunits.


\end{abstract}

\section{\starfire\ Instrument}

Understanding the formation and evolution of galaxies is one of the foremost goals of astrophysics and cosmology today. The cosmic star formation rate rose dramatically from early times to a peak at approximately half the present age of the universe (at redshift $z \sim 1$), with much of the activity occurring in highly dust-obscured systems. A variety of atomic and molecular diagnostic lines are present in the far-infrared (FIR) that are largely unaffected by dust, and can provide insight into the conditions of star formation at the cosmic peak. \starfire\ (the Spectroscopic Terahertz Airborne Receiver for Far-InfraRed Exploration) is designed to study the interstellar medium (ISM) of galaxies from $0.5 < z < 1.5$, primarily in the [C{\sc ii}] 158 \mum\ line, and also in cross-correlation with the [N{\sc ii}] 122 \mum\ transition. \starfire\ will be capable of making a high significance measurement of the \CII\ power spectrum in at least 4 redshift bins, and of measuring the \CII$\times$\NII\ power spectrum at $z \sim 1$ \citep{uzgil14}. \starfire\ will also be able to detect emission lines in a blind survey, and by correlating with known optical galaxies measure the \CII\ luminosity of this population as well. 
\starfire\ is able to achieve its substantial increase in performance -- better than the airborne instruments on SOFIA or the space-borne {\em Herschel}-SPIRE FTS -- by using dispersive spectroscopy to lower the photon noise per detector, and by taking advantage of the considerably lower atmospheric background at balloon rather than aircraft altitudes. \starfire\ will field two large format ($\sim$1800 detectors each) arrays of dual-polarization sensitive detectors with NEPs below the typical photon NEP of $1.5 \times 10^{-17}$ \nepunits. \starfire\ serves as a technology advancement platform for the Origins Space Telescope \citep{OSTwhitepaper}, and detector development is currently funded by NASA \citep{barlis17}.

\section{Detector Design and Fabrication}


\begin{SCfigure}[][t]
\centering
\includegraphics[%
  width=0.50\linewidth,
  keepaspectratio]{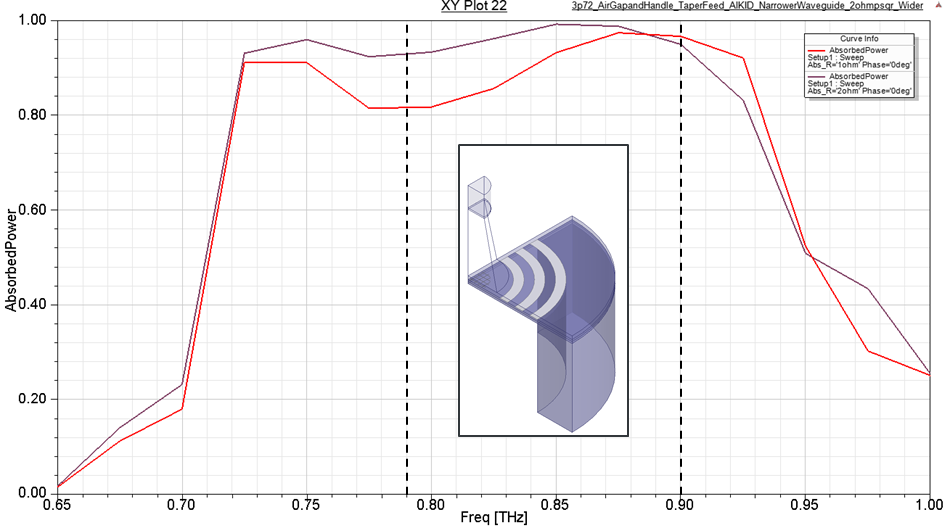}
\caption{\small Electromagnetic simulation of the coupling efficiency of a TE$_{11}$ mode from the circular waveguide into the aluminum absorber, for sheet impedances of 2 $\mathrm{\Omega}/\Box$ ({\it purple}) and 1 $\mathrm{\Omega}/\Box$ ({\it red}), and a 27 \mum\ backshort. Half power frequencies of the testbed bandpass filter are also shown ({\it dashed}). The band-averaged coupling efficiencies are 94\% and 88\% for 2 $\mathrm{\Omega}/\Box$ and 1 $\mathrm{\Omega}/\Box$ , respectively, and drop to 22\% and 20\% with no backshort. (Color figure online.)}
\label{fig:hfss_fig}
\end{SCfigure}

\starfire\ will deploy arrays of kinetic inductance detectors (KIDs) with the same single-layer architecture developed for the MAKO camera \citep{McKenney2012}. A single 20 nm thick aluminum layer forms both the inductor and interdigitated capacitor, which are designed to achieve $100-250$ MHz readout frequencies. To maximize responsivity the absorber is low volume ($V = 38$ \mum$^3$), and couples to incident radiation with a circular waveguide fed with a conical feedhorn. The waveguide design includes a flare at the bottom of the waveguide and a lithographically patterned choke structure to help eliminate conversion into substrate modes. The final pixel will be fabricated on a SOI wafer and will have a 27 \mum\ thick backshort, created by etching from the backside to a buried oxide layer, then depositing gold. Electromagnetic simulations indicate band-averaged coupling efficiencies of $\approx$$90\%$ with the backshort in place, and $\approx$$20\%$ without (Figure \ref{fig:hfss_fig}). Initial testing of a device with the backshort fabricated indicate the presence of the gold has no noticeable impact on the resonator $Q$. 

\begin{figure}[t]
\centering
  \includegraphics[width=\textwidth]{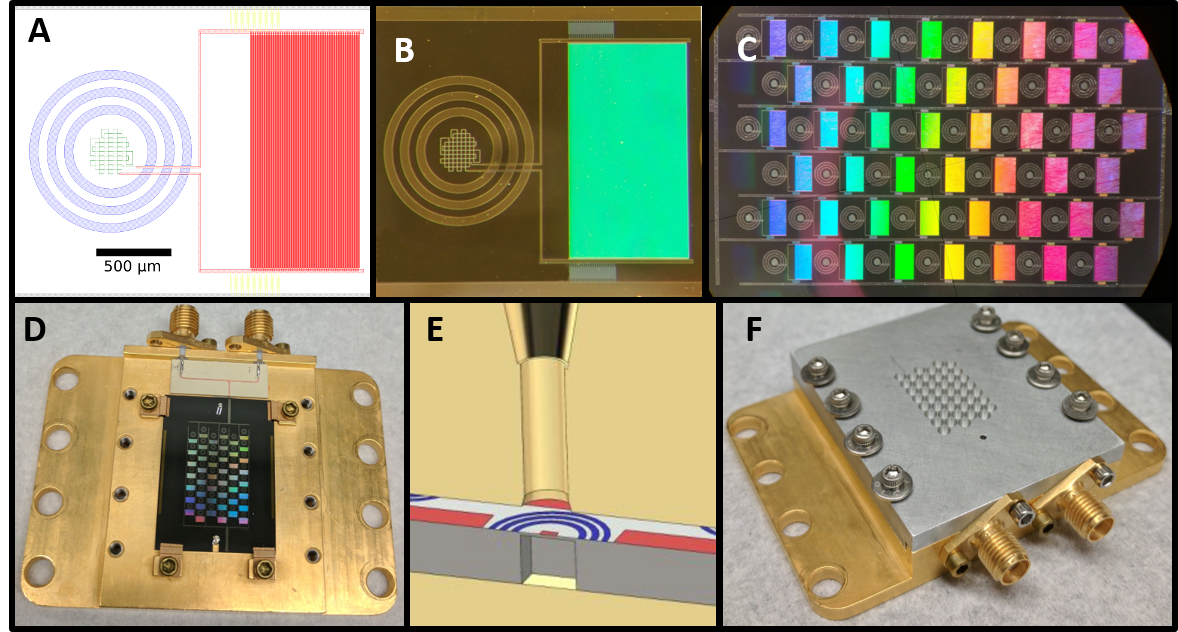}
  \caption{\small (A) Diagram of the mask layout for a single resonator. The meandered inductor ({\it green}) is surrounded by an optical choke structure ({\it blue}). An interdigitated capacitor ({\it red}) sets the resonance frequency of the pixel, and two coupling capacitors ({\it yellow}) couple microwave signal onto microstrip feedlines. (B) A microscope image of a single pixel as fabricated. All pixel elements of the prototype array are patterned out of 40 nm Al film. (C) A microscope image of the 45-pixel prototype array, as fabricated. (D) The fabricated array in its enclosure. The back side of the die is bare silicon, and lies flat on the gold-plated package surface. The full size of the die is 30mm $\times$ 22mm. (E) A CAD model of the detector package. The optical power is coupled into a feedhorn, and travels through a circular waveguide that is terminated by the inductor of the LEKID. The final design will have a backside etch providing a backshort. (F) The prototype feedhorn block installed above the 45-pixel array. (Color figure online.)}
\label{fig:pixel_fig}
\end{figure}

The inductor/absorber is a single meander of 0.4 \mum\ wide aluminum, patterned to provide an optimal impedance match to the waveguide. The meander effectively couples as a mesh to both polarizations by allowing the various segments of meander line to come close enough to one another at the corners to create capacitive shorts at the optical frequencies. This is achieved with a 0.3 \mum\ gap and a 0.6 \mum\ overlap length for each of the intersections (Figure \ref{fig:pixel_fig}).



\section{Device Testing}

We have cryogenically tested a 45-pixel prototype detector array, fabricated in the JPL Microdevices Laboratory. To measure the performance of detectors with different film thicknesses we fabricated wafers with both 20 nm and 40 nm thick aluminum, and conducted the initial testing using an array with a 40 nm thick film. This results in a larger inductor volume ($V = 76$ \mum$^3$) and higher resonance frequencies than expected for the full \starfire\ array, which will use a 20 nm thick film. Measurements of the sheet impedances of these films are in preparation. The array is cooled by a $^3$He sorption fridge to a base temperature of 210 mK, and exposed to a cryogenic blackbody for optical testing. We use two metal-mesh filters mounted on the detector package to define the optical band: a bandpass filter transmitting over $\approx 790 - 900$ GHz, and a 1000 GHz cutoff low-pass filter. We use a ROACH-based readout system originally developed for use with MAKO \citep{McKenney2012}. The prototype device had a total yield of 89\% (40/45 resonators), but we focus our analysis on the 11 resonators with resonances below 250 MHz.  

\subsection{Dark Measurements}

\begin{figure}[t]
\begin{center}
\includegraphics[width=0.96\linewidth, keepaspectratio]{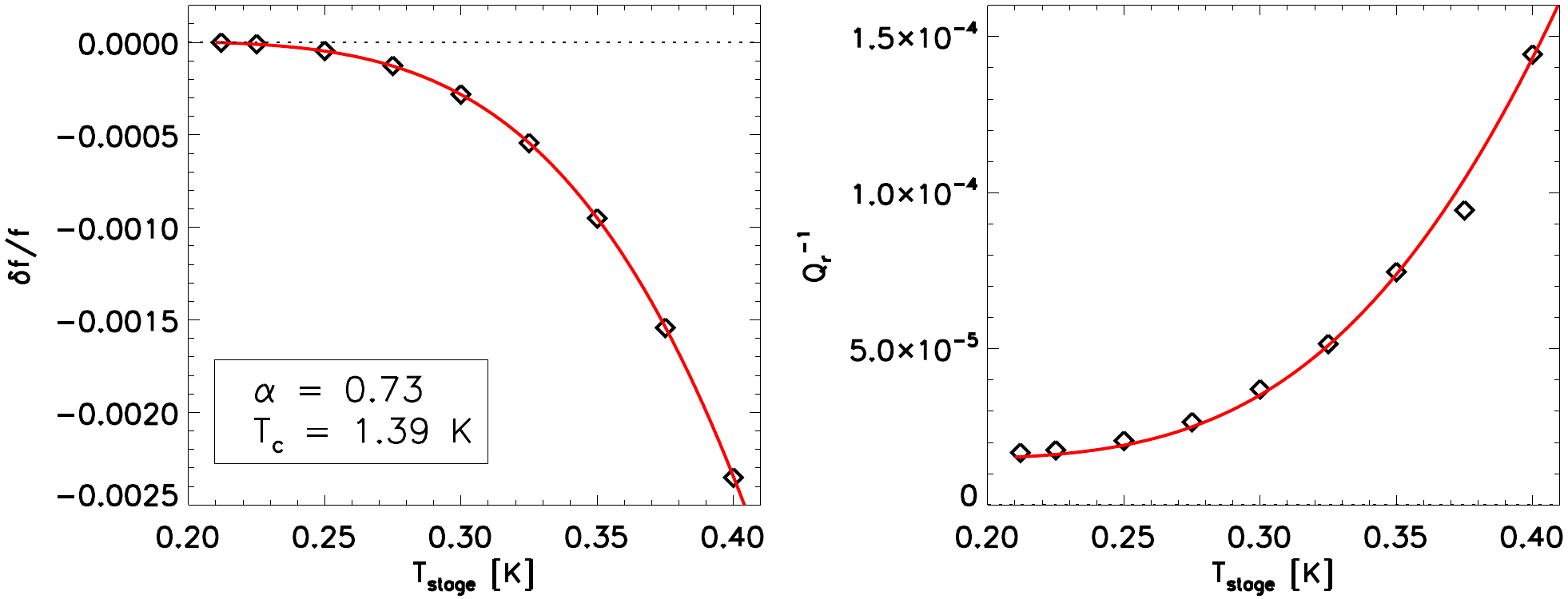}
\caption{\small Fractional frequency shift and $Q_r^{-1}$ vs. $T_\mathrm{stage}$ for a dark measurement, along with fits for $\alpha$ and $T_c$. Models are $x = x_\mathrm{MB} + \delta x$ for a zero temperature offset $\delta x$, and $Q_r^{-1} = Q_\mathrm{MB}^{-1} + Q_\mathrm{0}^{-1}$, with $Q_\mathrm{0}^{-1}$ a fixed term that includes the resonator coupling $Q$ and a limiting inductor internal $Q$. (Color figure online.)}
\label{fig:x_Q_dark}
\end{center}
\end{figure}

Our first step is to characterize the detectors dark, with the feedhorns blanked off. We measure the resonator frequency, $Q$, and noise as a function of stage temperature. We model our resonators following the standard application of Mattis-Bardeen theory, along with the assumption that the quasiparticle lifetime depends on the quasiparticle number density as $\tau_\mathrm{qp} = \tau_\mathrm{max} (1 + n_\mathrm{qp}/n^*)^{-1}$, for constants $\tau_\mathrm{max}$ and $n^*$ \citep{Gao2008JLTP,Zmuidzinas2012}. For a general temperature and absorbed optical power ($P_\mathrm{abs}$), $n_\mathrm{qp}$ may be written as:

\begin{equation}
n_\mathrm{qp} = -n^* + \bigg[(n^* + n_\mathrm{th})^2 + \frac{2n^* \eta_\mathrm{pb} P_\mathrm{abs} \tau_\mathrm{max}}{\Delta_0 V} \bigg]^{0.5},
\label{eq:nqp_general}
\end{equation}

\noindent where $n_\mathrm{th} = 2 N_0 \sqrt{2 \pi k_B T \Delta_0} \, \mathrm{exp}(-\Delta_0/k_B T)$ is the quasiparticle density in thermal equilibrium, $\eta_\mathrm{pb}$ is the pair-breaking efficiency, $\Delta_0 = 1.76k_BT_c$ is the gap energy, $V$ is the inductor volume, and we adopt a density of states of $N_0 = 1.72 \times 10^{10}$ $\mu$m$^{-3}$\,eV$^{-1}$ \citep{Gao2008JLTP}. The fractional frequency shift and the internal $Q$ of the inductor are then written as:

\begin{equation}
x_\mathrm{MB} = -\frac{\alpha \gamma S_2}{4 N_0 \Delta_0} n_\mathrm{qp} \quad\text{and}\quad Q_\mathrm{MB}^{-1} = \frac{\alpha \gamma S_1}{2 N_0 \Delta_0} n_\mathrm{qp},
\label{eq:x_mb}
\end{equation}

\noindent where $\gamma = 1$ is appropriate for the thin films used here. We use the standard expressions for $S_1$ and $S_2$ \citep{Zmuidzinas2012}, but in place of the physical temperature of our devices we substitute an effective electron temperature, obtained by inverting $n_\mathrm{th} (T)$ for $n_\mathrm{qp}$ \citep{Barry2017}. This modified approach becomes important when the absorbed power is nonzero, and $n_\mathrm{qp} > n_\mathrm{th}$. In Figure \ref{fig:x_Q_dark} we show measurements of $x$ and $Q_r^{-1}$ as a function of $T_\mathrm{stage}$ for one resonator, along with fits for $\alpha$ and $T_c$. The resonators are well-characterized by Equations \ref{eq:nqp_general} and \ref{eq:x_mb}, with median values of $\alpha = 0.74$ and $T_c = 1.39$ K.


\begin{figure}[t]
{\includegraphics[width=0.48\textwidth]{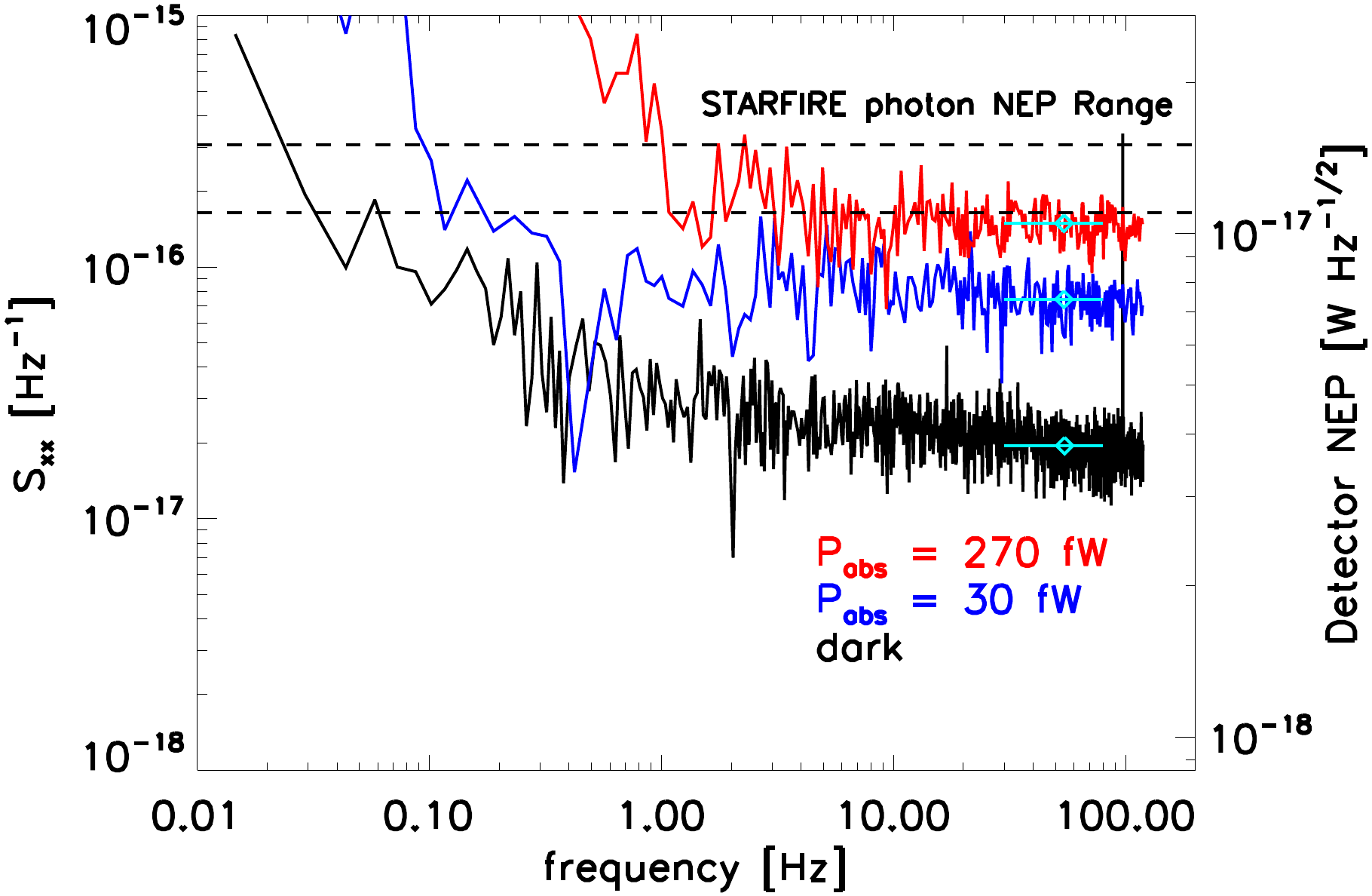}
}
\hfill
{\includegraphics[width=0.48\textwidth]{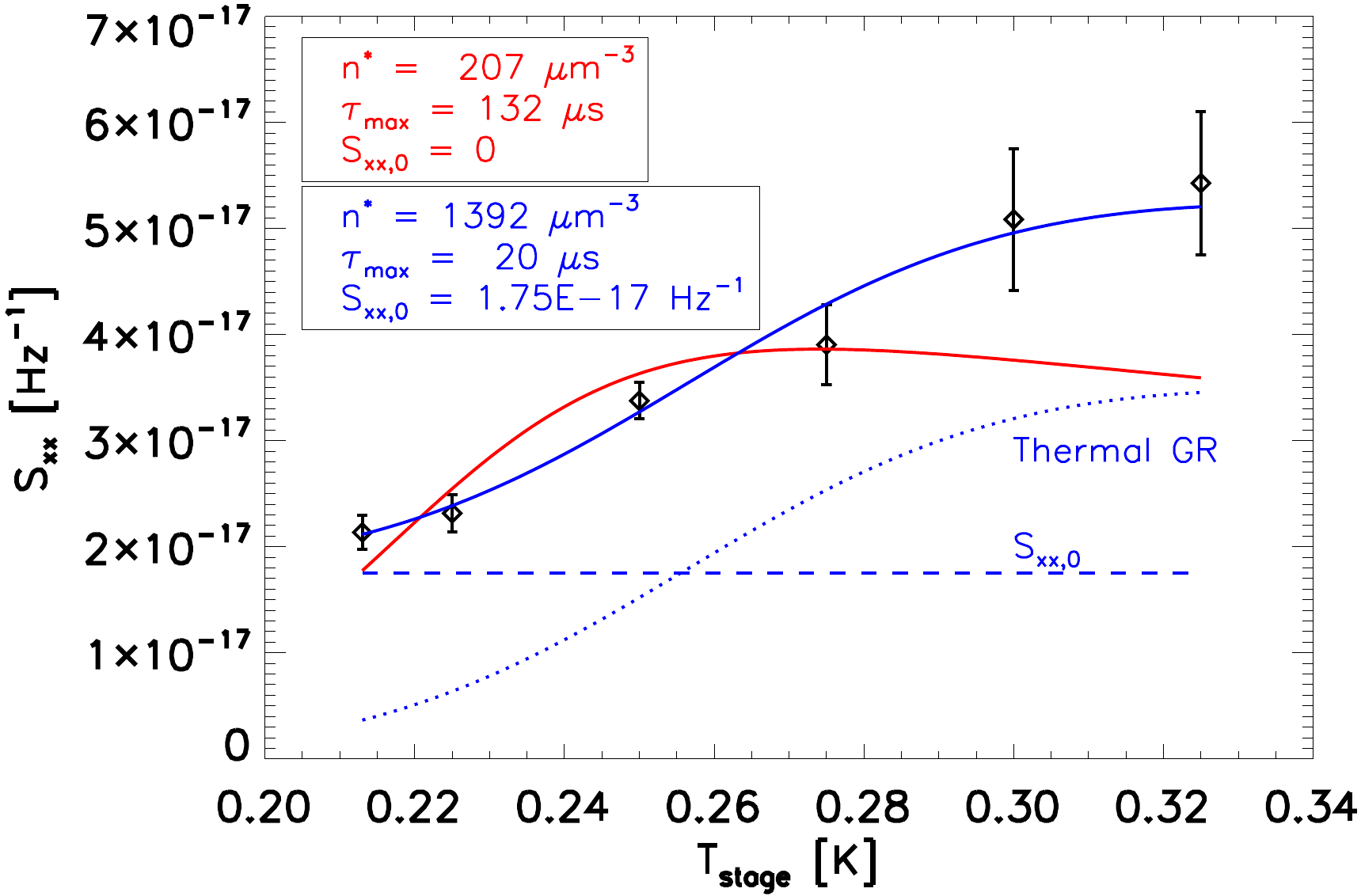}
}
\caption{\small {\it Left} Amplifier-subtracted $S_\mathrm{xx}$ for a representative KID measured dark ({\it black}), and with the cryogenic blackbody delivering $P_\mathrm{abs} = 30$ fW ({\it blue}) and 270 fW ({\it red}). White noise is estimated by averaging over $30 - 80$ Hz ({\it cyan}), and conversion to a detector NEP uses the responsivity $R = 1.2\times10^9$ W$^{-1}$ measured at the lowest optical loading $P_\mathrm{abs} = 20$ fW. {\it Right} Amplifier-subtracted $S\mathrm{xx}$ measured dark at various stage temperatures, averaged over 6 KIDs. Fits to Equation \ref{eq:sxx_dark} are shown with $S_{xx,0}=0$ ({\it red}) and $S_{xx,0}$ a free parameter ({\it blue solid}), with the latter decomposed into thermal GR ({\it blue dotted}) and fixed noise floor ({\it blue dashed}). (Color figure online.)}
\label{fig:nep_fig}
\end{figure}

We measure the fractional frequency noise ($S_\mathrm{xx}$) at each stage temperature, driving the resonators $\sim$3 dB below bifurcation to minimize amplifier noise. In all measurements presented here the amplifier noise is measured in the dissipation direction, and subtracted off from the measured $S_\mathrm{xx}$ (Figure \ref{fig:nep_fig}). Our general model for the white noise in our KIDs combines photon generation noise, thermal generation noise, and recombination (of all quasiparticles) noise, along with a fixed term ($S_\mathrm{xx,0}$) representing additional noise sources, assumed to be independent of temperature and optical loading:

\begin{equation}
S_{xx} = \bigg(\frac{\alpha \gamma S_2}{4 N_0 \Delta_0}\bigg)^2 \bigg[\bigg( \frac{\eta_\mathrm{pb} \tau_\mathrm{qp}}{\Delta_0 V} \bigg)^2 2h\nu P_\mathrm{abs}(1+n_\gamma) + \frac{4(\tau_\mathrm{qp})^2}{V^2} (\Gamma_\mathrm{th} + \Gamma_r ) \bigg] + S_{xx,0},
\label{eq:sxx_general}
\end{equation}

\noindent where $n_\gamma$ is the photon occupation number in the detector, the thermal generation rate is $\Gamma_\mathrm{th} = (n_\mathrm{th}V/2) (\tau_\mathrm{max}^{-1} + \tau_\mathrm{th}^{-1})$, $\tau_\mathrm{th}$ is the quasiparticle lifetime when $n_\mathrm{qp} = n_\mathrm{th}$, and the recombination rate is $\Gamma_r = (n_\mathrm{qp}V/2) (\tau_\mathrm{max}^{-1} + \tau_\mathrm{qp}^{-1})$ \citep{Zmuidzinas2012}. In the limit of no optical loading this becomes:

\begin{equation}
S_{xx} \rightarrow \bigg( \frac{\alpha \gamma S_2}{4 N_0 \Delta_0} \bigg)^2 \frac{4 n_\mathrm{th} \tau_\mathrm{th}}{V} \bigg(1 + \frac{\tau_\mathrm{th}}{\tau_\mathrm{max}}\bigg) + S_{xx,0}.
\label{eq:sxx_dark}
\end{equation}

\noindent The temperature dependence of Equation \ref{eq:sxx_dark} is dominated by the $n_\mathrm{th} \tau_\mathrm{th} (1 + \tau_\mathrm{th}/\tau_\mathrm{max})$ product. At low temperatures this term reduces to $2 n_\mathrm{th} \tau_\mathrm{max}$ and increases rapidly with temperature, while at high temperatures this term asymptotes to $n^* \tau_\mathrm{max}$ \citep{Mauskopf2014}. In Figure \ref{fig:nep_fig} we show the white noise as a function of temperature, averaged over a subset (6/11) of the KIDs with the lowest amplifier noise. We show a fit to the data with fixed $P_\mathrm{abs} = S_{xx,0} = 0$, and $n^*$ and $\tau_\mathrm{max}$ as free parameters. This fit is poor, but the data are well reproduced by introducing a noise floor $S_{xx,0} \approx 1.8 \times 10^{-17}$ Hz$^{-1}$, or alternatively by assuming a frequency-dependent stray power $P_\mathrm{abs} \approx 20\,(100\,\mathrm{GHz}/\nu)$ fW for a minimum pair-breaking frequency of 100 GHz. These fits indicate thermal generation-recombination (GR) noise is subdominant to other noise sources at our 210 mK operating temperature, but dominates at $T > 250$ mK.

\subsection{Optical Measurements}

\begin{figure}[t]
\begin{center}
\includegraphics[width=0.96\linewidth, keepaspectratio]{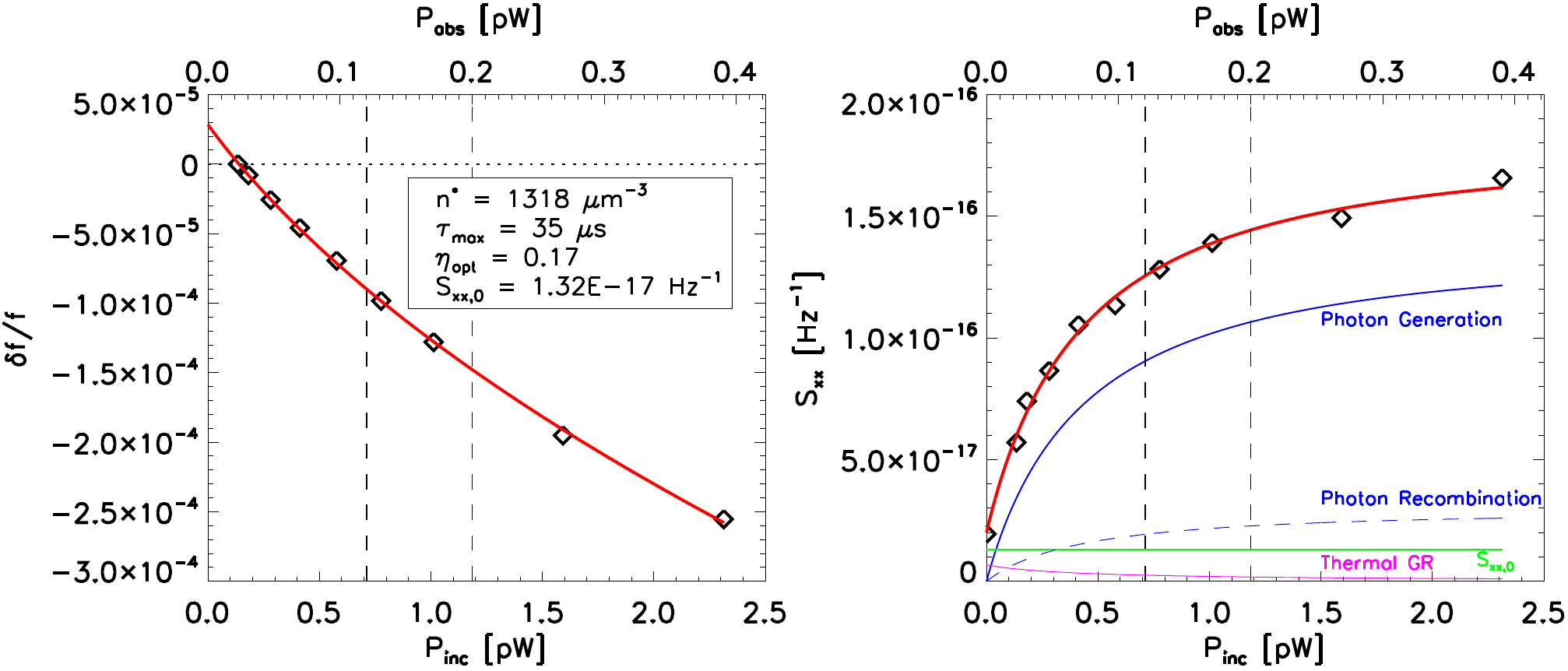}
\caption{\small {\it Left} Fractional frequency shift vs. $P_\mathrm{inc}$ with model fit ({\it red}). {\it Right} $S_\mathrm{xx}$ vs. $P_\mathrm{inc}$ with model fit ({\it red}), decomposed into contributions from photon generation noise ({\it blue solid}), photon recombination noise ({\it blue dashed}), thermal GR noise ({\it magenta}), and fixed noise floor ({\it green}). Vertical dashed lines in both figures mark the minimum and median \starfire\ optical loading of 120 fW and 200 fW, respectively. (Color figure online.)}
\label{fig:x_sxx}
\end{center}
\end{figure}

In a second test we expose the detector package to the cryogenic blackbody, and measure the frequency response and noise as a function of the blackbody temperature for $T_\mathrm{BB} = 6.2 - 11$ K. We compute the power incident on the feedhorn aperture by integrating the blackbody source function over the filter transmission profiles, assuming a single spatial mode and dual polarization response. Figure \ref{fig:x_sxx} shows the fractional frequency response and white noise for a representative channel. The slope of the response curve is the product of the responsivity and optical efficiency ($\eta_\mathrm{opt}$):

\begin{equation}
\frac{\delta x}{\delta P_\mathrm{inc}} = \frac{\delta x}{\delta P_\mathrm{abs}} \eta_\mathrm{opt} = \frac{\alpha \gamma S_2}{4 N_0 \Delta_0} \frac{\eta_\mathrm{pb} \tau_\mathrm{qp}}{\Delta_0 V} \eta_\mathrm{opt},
\end{equation}

\noindent and the flattening of this curve at high $P_\mathrm{inc}$ is consistent with the expected decrease in $\tau_\mathrm{qp}$ at large $n_\mathrm{qp}$. 

We see an increase in white noise with $P_\mathrm{inc}$ that we attribute to photon noise (Figure \ref{fig:nep_fig}). The contribution of photon noise to $S_\mathrm{xx}$ is:

\begin{equation}
S_\mathrm{xx,\gamma} = \bigg(\frac{\delta x}{\delta P_\mathrm{inc}}\bigg)^2 \frac{2h\nu P_\mathrm{inc}}{\eta_\mathrm{opt}} \bigg(1 + \frac{2\Delta_0}{h\nu\eta_\mathrm{pb}}\bigg),
\label{eq:sxx_gamma}
\end{equation}

\noindent where we have neglected wave noise (negligible for $T_\mathrm{BB} \le 11$ K), and the last term accounts for the recombination noise associated with optically-generated quasiparticles. With an assumption of $\eta_\mathrm{pb} = 0.57$ this term is 1.21, and Equation \ref{eq:sxx_gamma} then demonstrates how the ratio of $S_\mathrm{xx,\gamma}$ and $(\delta x/\delta P_\mathrm{inc})^2 P_\mathrm{inc}$ may be used to estimate $\eta_\mathrm{opt}$. 

We fit the response and noise data shown in Figure \ref{fig:x_sxx} using our full resonator model (Equations \ref{eq:nqp_general}$-$\ref{eq:sxx_general}), with free parameters $n^*$, $\tau_\mathrm{max}$, $\eta_\mathrm{opt}$, and $S_\mathrm{xx,0}$. The data are well fit with this model, and we find median values of $n^* = 1240$ $\mu$m$^{-3}$, $\tau_\mathrm{max} = 35$ $\mu$s, $\eta_\mathrm{opt}=0.17$, and $S_{xx,0} = 1.2\times10^{-17}$ Hz$^{-1}$. This value of $\tau_\mathrm{max}$ is shorter than reported by other groups for aluminum \citep{McCarrick2014,Baselmans2017}. Our aluminum is currently deposited through a sputter deposition technique; a future shift to electron beam evaporation may produce higher purity films and longer quasiparticle lifetime, increasing the responsivity \cite{barends09}. This optical efficiency is close to the $0.20-0.22$ range estimated from simulations of the coupling efficiency between the waveguide and absorber (Figure \ref{fig:hfss_fig}). The inferred responsivity at our lowest optical loading ($P_\mathrm{abs} = 20$ fW) averaged over the array is $\delta x/\delta P_\mathrm{abs} = 1.2\times10^9$ W$^{-1}$. We combine this with our dark noise measurements to obtain detector NEPs, finding a median value of $4\times10^{-18}$ \nepunits\ (Figure \ref{fig:nep_fig}).

The \starfire\ arrays will operate with a typical optical loading of 200 fW, at a 250 mK base temperature. Increasing the optical load from 20 fW to 200 fW decreases the responsivity in our prototype KIDs by $\approx$$40\%$ (Figure \ref{fig:x_sxx}). Additionally, the dark noise at 250 mK is $\approx$$1.5$ times larger than at 210 mK (Figure \ref{fig:nep_fig}). Combined, this indicates an NEP of $8 \times 10^{-18}$ \nepunits\ with the optical load and operating temperature envisioned for \starfire, less than the photon NEP of $1.5 \times 10^{-17}$ \nepunits.


\section{Summary}

We have fabricated and characterized a 45-pixel \starfire\ prototype detector array. These LEKIDs are low volume (76 \mum$^3$) devices fabricated with a single layer of 40 nm thick aluminum, are sensitive to both polarizations, and couple to free space with circular waveguide and conical feedhorns. Operating at 210 mK we measure a typical NEP of $4\times10^{-18}$ \nepunits, and confirm that thermal GR noise is not the dominant noise source. With a 250 mK operating temperature and under a 200 fW load, as we anticipate for \starfire, the detector NEP will be $\approx 8 \times 10^{-18}$ \nepunits. This compares favorably to the typical photon NEP of $1.5 \times 10^{-17}$ \nepunits.





\begin{acknowledgements}
ACMB's work was supported by a NASA Space Technology Research Fellowship. Detector development for \starfire\ is supported by NASA grant 15-APRA15-0081. We thank C. Groppi for generously providing the tool used to drill the feedhorns.
\end{acknowledgements}


\bibliography{List} 

\end{document}